\documentclass{aa}
\usepackage{psfig}
\usepackage{graphics}
\def\utw{\smash{\rlap{\lower5pt\hbox{$\sim$}}}}
\def\udtw{\smash{\rlap{\lower6pt\hbox{$\approx$}}}}

\def\diameter{{\ifmmode\mathchoice
{\ooalign{\hfil\hbox{$\displaystyle/$}\hfil\crcr
{\hbox{$\displaystyle\mathchar"20D$}}}}
{\ooalign{\hfil\hbox{$\textstyle/$}\hfil\crcr
{\hbox{$\textstyle\mathchar"20D$}}}}
{\ooalign{\hfil\hbox{$\scriptstyle/$}\hfil\crcr
{\hbox{$\scriptstyle\mathchar"20D$}}}}
{\ooalign{\hfil\hbox{$\scriptscriptstyle/$}\hfil\crcr
{\hbox{$\scriptscriptstyle\mathchar"20D$}}}}
\else{\ooalign{\hfil/\hfil\crcr\mathhexbox20D}}%
\fi}}
\def\squareforqed{\hbox{\rlap{$\sqcap$}$\sqcup$}}
\def\sq{\ifmmode\squareforqed\else{\unskip\nobreak\hfil
\penalty50\hskip1em\null\nobreak\hfil\squareforqed
\parfillskip=0pt\finalhyphendemerits=0\endgraf}\fi}

\makeatletter

\def\@biblabel#1{}
\renewenvironment{thebibliography}[1]
     {\section*{\refname
        \@mkboth{\refname}{\refname}}\small
      \list{\@biblabel{\@arabic\c@enumiv}}%
           {\settowidth\labelwidth{\@biblabel{#1}}%
           \leftmargin\bibindent
           \parsep\z@\itemsep\z@
           \setlength{\itemindent}{-\leftmargin}
           \@openbib@code
           \usecounter{enumiv}%
           \let\p@enumiv\@empty
           \renewcommand\theenumiv{\@arabic\c@enumiv}}%
      \sloppy\clubpenalty4000\widowpenalty4000%
      \sfcode`\.\@m}
     {\def\@noitemerr
       {\@latex@warning{Empty `thebibliography' environment}}%
      \endlist}
\renewcommand\newblock{\hskip .11em\@plus.33em\@minus.07em}
\let\@openbib@code\@empty

\makeatother

\begin{document}

\thesaurus{03
(11.07.1;    % Galaxies: general
11.09.4;    % Galaxies: ISM
11.10.1;    % Galaxies: jets
11.11.1;    % Galaxies: kinematics and dynamics
13.18.1;    % Radio continuum: galaxies
13.19.3% Radio lines: ISM
)
}

\title{Radio Jet-blown Neutral Hydrogen Supershells in Spiral Galaxies?}

\titlerunning{Jet-blown HI supershells in spirals}

\author{Gopal-Krishna\inst{1} \and Judith.A. Irwin\inst{2}} 

\authorrunning{Gopal-Krishna \& J. A. Irwin}

\institute{National Centre for Radio Astrophysics, Tata Institute 
of Fundamental Research, Pune University Campus, Pune - 411007, India 
\thanks{e-mail: krishna@ncra.tifr.res.in} \and Dept. of Physics, Queen's 
University, Kingston, Canada, K7L 3N6\thanks{e-mail:
irwin@astro.queensu.ca}}

\date{}

\maketitle{}

\begin{abstract}
Taking a clue from the pair of HI supershells found in the Scd
galaxy NGC~3556 (M~108), we propose a new mechanism for the origin
of HI supershells in gas-rich massive galaxies. 
In this scenario, the two supershells
 were inflated out of the neutral hydrogen disk
due to the localised flaring of a pair of radio lobes formed by the
jets ejected from
the nucleus during an active phase about $\sim~10^7$ years ago, but
have faded away by now. It is shown that the salient features of this
supershell pair, such as their symmetrical locations about the galactic
centre, the anomalously large energy requirements, the large
galacto-centric distances, as well as the Z-symmetric hemispherical
shapes, find a more natural explanation in terms of this scenario, as
compared to the standard models which postulate either a massive
starburst, or the infall of external gas clouds. Other possible
implications of this hypothesis are briefly discussed.
\end{abstract}

\keywords{
Galaxies: general --
Galaxies: ISM --
Galaxies: jets --
Galaxies: kinematics and dynamics --
Radio continuum: galaxies --
Radio lines: ISM --
}

\section{Introduction}

An important manifestation of the activity inside the disks of
gas-rich galaxies is their highly structured HI distribution, marked
by cavities, shells and supershells. First discovered in the Milky Way
(Heiles 1979, 1984), such features are now known to exist in a number
of spiral galaxies (e.g. Lehnert \& Heckman 1996, Irwin \& Seaquist 1990,
Puche et al. 1992, Brinks \& Bajaja 1986).
%Gopal there are 2 references to Lehnert & Heckman 1996 -- please check that
%I have the correct one.
Exceptionally huge HI arcs and loops extending across several kiloparsecs
have been identified with greater clarity in the HI images of a number
of edge-on spirals, such as NGC~5775 (Irwin 1994), NGC~4631 (Rand \& van
der Hulst 1993), NGC~3044 (Lee \& Irwin 1997, hereafter LI97) and NGC~3556
(M 108, King \& Irwin 1997, hereafter KI97).  These have been interpreted
as expanding supershells because of a loop-like or circular appearance in
projection and either a persistence over a wide velocity range or, in a
few cases, as some evidence for expansion in position-velocity
space. Two main classes of explanations for the supershells posit the source
of their kinetic energy to be, respectively, internal and external to
the parent galaxy.

The internal source model involves starbursts, driving
stellar winds (or superwinds) and subsequent supernova explosions (e.g.
Lehnert \& Heckman 1996).  The ``chimney model" (Norman \& Ikeuchi 1989),
for example, attempts to explain disk-halo features and
other halo gas via processes related to underlying star formation.
The association between extraplanar H$\alpha$ filaments and star forming
regions in the disk of NGC~891 and other correlations between halo emission
and in-disk tracers of star formation (Dahlem et al. 1995;
Rand 1997) argue in favour of such models.  If the presence of HI supershells
is found to
correlate with the existence of other halo gas, as might be expected in
the chimney model, then stellar winds and supernovae are expected to be
responsible for the HI supershells as well.

The main difficulty with the starburst model for HI supershells lies in
the required input energies for the largest shells.  Using standard
assumptions that the expanding supershells are in the post-Sedov phase
following an `instantaneous' injection of energy (cf. Chevalier 1974), HI
supershells often require energy input from staggering numbers of spatially
correlated supernova events.  This was realized early on for our own
Galaxy (Heiles 1979, 1984).  For external edge-on galaxies, since we
are selectively observing only the largest shells, the energy deficit
problem is exacerbated.  In some cases, hundreds of thousands of
clustered supernovae are required (e.g. KI97, LI97), a conclusion which
is not changed significantly if the energy is injected continuously
over the lifetime of the shells.   Other evidence against star formation
processes creating the HI shells is also emerging.  Rhode et al. (1999)
find no optical evidence for recent star formation in the numerous
lower energy HI holes of Holmberg II and note that X-ray and FUV emission
are also absent.  They conclude that supernovae have not played a part in
the formation of the HI shells.  Efremov et al. (1998) outline numerous
other examples in which there appears to be no relation between HI shells
and star formation. They, as well as
Loeb \& Perna (1998),  propose that the HI shells are produced, instead,
by gamma ray bursts.

The alternative external source hypothesis invokes infall of massive gas
clouds on to the galactic plane, as a result of gravitational interaction
with neighbouring galaxies (see Tenorio-Tagle \& Bodenheimer 1988).  This
resolves the energy problem since input energy is then a function of the
mass and velocity of the infalling cloud.  Evidence in favour of this
hypothesis comes from observations of high velocity clouds (HVCs) around
our own Milky Way and the signatures of interaction in M~101 (van der Hulst
\& Sancisi 1988) and NGC~4631 (Rand \& Stone 1996).  It does, however,
require that the galaxy be in some way interacting with a companion or,
at least, that sufficiently massive clouds be in the vicinity.

Recent observations are revealing galaxies which are apparently isolated,
yet harbour extremely large HI supershells.  Two striking examples are
the nearby, SB(s)cd galaxy, NGC~3556 (KI97) and the SBc galaxy, NGC~3044
(LI97).  Both of these galaxies exhibit radio continuum halos extending
to $\sim~5$ kpc from the galactic plane and have a number of
supershells requiring energies up to a few $\times$ 10$^{56}$ ergs.
These supershells are too large and energetic to have been produced by
conventional clustered supernovae.  At the same time, there appears to be
 no evidence for interaction or nearby companions, either.

We propose here a new explanation for HI supershells. That is, that they
have been formed by radio jets which plow through the interstellar medium
(ISM), accreting ISM gas and sometimes inflating bubbles.  This allows for
an internal energy source for the HI shells, provides a natural explanation
for any spatial symmetries seen in the HI features, and also resolves the energy
problem. In Sect.~2, we provide arguments in favour of jet inflated HI bubbles,
Sect.~3 presents the model, and Sect.~4 discusses the implications of this scenario.

\section{The Case for a Radio Jet Origin for HI Supershells}

\subsection{Radio Jets and AGN in Spiral Galaxies}

Seyferts are one class of disk galaxy for which several examples of
the nucleus ejecting a radio jet pair have been found (e.g. Ulvestad
\& Wilson 1984a, 1984b, Kukula et al. 1995, Aoki et al. 1999).
Likewise, several cases of jets occurring in normal spiral galaxies
have been reported (e.g. Hummel et al. 1983).  Prominent
examples include NGC~3079 (de Bruyn 1977), NGC~5548 (Ulvestad et al. 1999)
and Circinus (Elmouttie et al. 1998).

The total energy output from such nuclear activity can approach $10^{60}$
erg, assuming that the nuclear activity lasts $\sim~10^7$ years, and
 arises from accretion onto a supermassive black hole of $10^8 M_{\odot}$
at 10\% of the Eddington limit with the canonical 10\% efficiency.  In
``normal" spiral galaxies, the central mass may be lower;  for example,
the bipolar outflow from the SAB(rs)cd galaxy, M~101  may be produced by
a central $\sim$ 10$^6$ M$_\odot$ black hole (Moody et al. 1995).  While
jets are not always observed directly, the growing number of supermassive
black holes inferred to be present in the nuclei of normal disk galaxies
(Kormendy \& Richstone 1995, van der Marel 1999), including late-type disks
(Ho et al. 1997), the Milky Way itself (e.g. Genzel et al.
1997) and gas-rich, large low-surface-brightness galaxies (Schombert 1998),
lend weight to the idea that many such galaxies may have undergone phases
of nuclear activity accompanied with a bi-directional ejection of
relativistic plasma jets, before entering a dormant nuclear phase.

Recent studies have also revealed that the ejection of jets can take place
at small angles to the large-scale galactic disk  (e.g. 
Nagar \& Wilson 1999; Kinney et al 2000),
plausibly leading to clear signatures of their dynamical interaction with
the ISM. A classic case of such an interaction is the bow-shaped morphology
of the radio lobe in the well known Seyfert galaxy, NGC~1068, which strongly
suggests that the kiloparsec-scale radio jets are ploughing through the disk
medium (Wilson \& Ulvestad 1987, Axon et al. 1998). A dynamical interaction 
of the jets  with the disk is also evident in the case of the spiral galaxy 
M~51 which is at
the low end of nuclear activity scale. Here, the jets seem to have created
a pair of `radio lobes' on opposite sides of the nucleus; whereas one of
the lobes is bow-shaped, the other one is actually resolved into a ring
which is also detected in H$\alpha$ emission (Ford et al. 1985).
Another example is NGC~4258 whose VLBI jet has been ejected close to
the galaxy disk (see Cecil et al. 1995,  Herrnstein et al. 1997,
Cecil et al. 1998).  Large scale (14 kpc length) ``anomalous arms" are also
seen {\it within} the galactic disk and have been interpreted as
manifestations of a larger-scale jet (e.g. Martin et al. 1989, but
see Cox \& Downes 1996).  In M~101, the bipolar outflow is also roughly
confined to the disk, with the outflow not expected to extend beyond a
height of 400 pc (Moody et al. 1995).  Additional support for the jet-disk
interaction hypothesis in these disk galaxies comes from the detection of
shock-excited optical emission lines associated with the radio lobes, as
discussed in some of the references cited above. 

\begin{figure*}[t]
%\psfig{figure=figure1.ps,width=5in}
%\resizebox{6in}{!}{\includegraphics{figure1.eps}}%,width=5in}
%\resizebox{4in}{!}{\includegraphics{figure1.eps}}%,width=5in}
%%%%%\resizebox{3.75in}{!}{\includegraphics{figure1.eps}}
%\resizebox{\hsize}{!}{\includegraphics{=figure1.ps}}
%,width=0.5 \textwidth,angle=0}
%%%%%%\vskip -0.8truein
\caption[]{ 
{\bf a.} Superimposition of two channels of HI emission of NGC~3556 from
 KI97 at velocities 633 km s$^{-1}$ (east side) and 757 
km s$^{-1}$ (west side).  The greyscale is set to emphasize the
HI loops on either side of the galaxy and ranges from 
0.9 mJy beam$^{-1}$ (2$\sigma$) 
to 3 mJy beam$^{-1}$.  Contours are at
0.9, 1, 1.25, 1.7, 
and 3 mJy beam$^{-1}$ and
 a white cross marks the galaxy's optical centre.
{\bf Inset.}  
Total intensity HI image of NGC~3556 (contours) from
 KI97 superimposed on an optical image of the galaxy.
{\bf b.} HI contours as in (a) 
superimposed on a greyscale 20 cm 
continuum map
(also from KI97) to emphasize the faint continuum features.
The greyscale ranges from 0.23 mJy beam$^{-1}$ (1$\sigma$)
to 1 mJy beam$^{-1}$. }
\end{figure*}

\subsection{Jet Inflated Bubbles in the Lobes of Radio Galaxies}

It is further interesting to note that radio bubbles and shells of
synchrotron plasma have been discovered within the lobes of a few radio
galaxies. Two spectacular examples are 3C310 (van Breugel \& Fomalont,
1984) and Hercules A (Dreher \& Feigelson 1984).  In Hercules A, in
addition, peculiar dark shells, about 3 kpc in size, have been found
straddling the nucleus along the radio outflow axes (Baum et al. 1996).
For the dark shells, Baum et al. prefer an explanation in terms of 
expanding bubbles of hot gas ejected from the active nucleus along the
radio axis, in agreement with plasmon-like ejection from the core, though 
they do not rule out other possibilities. 
% Very recently, another detection of a radio `bubble' has been reported in
%the northern radio lobe of Centaurus A (Morganti et al. 1999).

These cases provide a clue that the radio bubbles/shells may have 
been `puffed' up due to localized instabilities in the radio jets, occurring
at distances of a few kiloparsecs from the active nucleus (e.g. Morrison
\& Sadun 1996). Should such instabilities arise also
in the jets ejected within disk galaxies (Sect.~2.1), it is quite plausible
that the resulting increase in pressure at those locations would inflate
bubbles and shells out of the ambient ISM material which is HI rich,
in the case of spirals. Such features, puffed out of the disk during the
brief phase of jet instability, would remain visible past the radiative
lifetime of the synchrotron plasma against the radiative and expansion
losses, which is typically of order $10^6 - 10^7$ years.

\subsection{Candidate Galaxies}

\subsubsection{NGC~3556}

In NGC~3556, the two most prominent supershells are located symmetrically
near the opposite extremities of the galaxy.  KI97 have carried out a
detailed mapping of the HI and radio continuum emission from this nearly
edge-on galaxy located 11.6 Mpc away. 
The superimposition of 2 channels of HI emission, symmetrically placed
in velocity about systemic, are shown in Fig. 1 a.
The most obvious features are two loops of HI, producing
extensions to the east and west along the major axis.
Fig. 1b shows the HI emission from these velocities
superimposed on the radio continuum map, the latter
taken from entirely independent
observations.    The abrupt density drop-off in the
radio continuum map occurs roughly where the optical
galaxy ends (Fig. 1a inset).  The salient features of this
system most relevant to the present study (see Fig. 1; KI97) are:

%Gopal, I have taken out the part about the higher luminosity, including FIR.
% In fact, the  galaxy is only marginally brighter than the norm and if it
%*were* much brighter, this would argue for enhanced star formation which
%is not what we're claiming.

%Gopal, again the HI radius of 1.6 times the optical radius is pretty close to
%normal -- maybe a bit larger, cf. 1.4 times

(a) Two symmetrically placed giant HI loops are seen originating at a
projected galactocentric radius of about 12 kpc on the east and west major
axis.  They appear to originate right at the opposite edges of the optical
disk and extend, not perpendicular to the disk plane, but rather parallel
to it and outwards.

(b)  The giant HI loops or supershells are slightly bent into a 
Z-symmetry.

(c)  These HI loops are most obvious in velocity channels which are equally
spaced about the systemic velocity of the galaxy (though they extend over
a larger velocity range, see KI97).

(d) From their velocity dependent morphology, each of the loops shows
evidence for shell expansion, but with only half of the shell present.
In both cases, the receding side (with respect to galaxy rotation) is open.

(e)  Other smaller HI loops and extensions do exist at smaller galactocentric
radii. 

(f) The parameters of these two HI supershells are as follows: The eastern
supershell has a mass, M = 5.9 $\times$ 10$^7$ M$_\odot$, a radius, R = 3.2 kpc,
an expansion velocity, $\Delta v$ = 51.7 km s$^{-1}$, a kinematic age,
$\tau$ = 6.0 $\times$ 10$^7$ yr, a kinetic energy, E$_\mathrm{kin}$ = 1.6 $\times$
10$^{54}$ ergs, and an implied input energy (assuming instantaneous input
from supernovae) of E = 2.6 $\times$ 10$^{56}$ ergs.  The parameters of the
western supershell are:  M = 2.3 $\times$ 10$^7$ M$_\odot$,  R = 1.85 kpc,
$\Delta v$ = 41.4 km s$^{-1}$,  $\tau$ = 4.4 $\times$ 10$^7$ yr,
E$_\mathrm{kin}$ = 3.9 $\times$ 10$^{53}$ ergs, and  E = 3.4 $\times$ 10$^{55}$
ergs.

(g)  The energy needed to create the eastern supershell alone would require
a star cluster populated by $>~1.7 \times 10^5$ OB stars, if supernovae and
stellar winds are the drivers.

(h) The brighter and more spectacular eastern supershell has associated
HI gas which extends as far as 30 kpc beyond the eastern edge of the
optical disk and possibly much farther (see  Fig. 1a inset). 
The huge eastern
HI extension appears to open up from one side of the eastern shell and
extends fairly straight outwards in the radial, rather than the vertical
direction as if formed in a short period of time in comparison to galactic
rotation.

(i) The galaxy contains a large nonthermal radio halo with a scale height
of $\sim~5$ kpc above the plane (see also de Bruyn \& Hummel 1979, Bloemen
et al. 1993).

(j)  Radio continuum emission appears to be associated with the HI loops,
but is not spatially coincident with them.  The radio continuum emission
extends farther out than the HI supershells, as if ``funnelling" along or
through  the HI features (KI97; Fig. 1b).

(k)  NGC~3556 has no obvious interacting companions.  The brightest galaxy in the vicinity
is the 16th magnitude, MCG+09-19-001, 25$^{\prime\prime}$ in size, undisturbed in
appearance and 
$\sim$ 12$^{\prime}$ to the east of NGC~3556.  This is most likely a background
source.

\subsubsection{Other Examples}

NGC~5775, an interacting galaxy, shows symmetrically placed HI features
(Lee et al. 2000) in the sense that HI extensions occur on opposite sides
of the major axis in which there appears to be an underlying disturbance;
these features occur at similar galactocentric radii,
 in projection (see also Fig. 3 of Irwin 1994 which
shows 3 of the extensions).  NGC~2613 (Chaves \& Irwin 2000) also shows 
six HI extensions which occur in pairs symmetrically on either side of the major axis;
two of the pairs occur near the ends of the optical disk.

Another potentially very interesting example is M~31.  Blitz et al. (1999) show that
two massive HI clouds exist symmetrically placed on opposite sides of this
galaxy.  They bear a remarkable resemblance to extragalactic radio jets.
The two clouds are redshifted by $\sim~200$ km s$^{-1}$ with respect to the
systemic velocity of M~31. Conceivably, this could be explained by ram
pressure sweeping as M~31 falls through the IGM towards the Milky Way.

%Gopal, I have actually done these calculations but felt it might be
%better to leave them out of this paper.  This means that Fig. 2
%shouldalso be left out.  If this
%paper is well received, we could look at M31 in more detail later.

\section{A New Model for the Formation of HI Supershells}

\subsection{Need for a New Mechanism}

As mentioned above, the galaxy NGC~3556 puts to a severe test both
of the standard explanations for the supershell phenomenon, namely:
(i) localized starburst (generating intense stellar winds, followed
by multiple supernova explosions), and (ii) infall of massive gas
clouds on to the galactic disk.  The major difficulties faced by these
models, as highlighted in KI97, are:

(a) {\bf Localized starbursts:} The energy needed to create the eastern
supershell alone would require a star cluster populated by $>~1.7 \times
10^5$ OB stars. Even for the recently discovered `super star clusters (SSC)'
in some starburst galaxies, not more than $10^4$ OB stars are implied
(Meurer et al. 1995). There is no indication of an OB association even
remotely approaching the level of SSCs at the locations of the supershells
in NGC~3556.  Since the kinematic ages of the supershells are $\sim$ 5
$\times$ 10$^7$ yr, the starburst would have to have occurred within this
time frame.  Yet starburst durations and OB association ages are typically
also of this order, so some evidence of the starburst might be expected to
have survived.  No such evidence is presently found, however.  
NGC~3556 does not
show a markedly high supernova rate globally (Irwin et al. 1999) and
independent low and high resolution radio continuum observations show no
evidence for a source of energy at the bases of the supershells (KI97, 
Irwin et al. 2000).

(b) {\bf Infalling clouds:} Several arguments have been advanced 
against this possibility (KI97). 
Firstly, this galaxy appears isolated,
making the source of the putative clouds puzzlesome.  Secondly, to impart
sufficient kinetic energy for the creation of the eastern supershell, an
infalling cloud of mass $>~10^8~M_{\odot}$ would be required.  
A cloud this massive would easily have been detected on the
sensitive HI maps, yet no such cloud or clouds are seen. (An exception
would be if the clouds had very narrow line widths, low fractal dimensions
and were optically thick, in which case they could have been missed due
to low filling factors.)  Thirdly, if an encounter with intergalactic clouds
has occurred in the past,
it would be necessary to postulate that
two very massive clouds just happened to fall in parallel to the major
axis at opposite ends of the galaxy, roughly at the same time, a
scenario which seems implausible. Fourthly, we could instead suggest that
a ``rain" of HI clouds is infalling (since other high latitude HI
structures are also seen in this galaxy), 
including two which fell in along the
major axis at either end.  However, in order for the rain to be no
longer visible, 
the infall would have to be completed over a timescale
which is less than or equal to the age of the supershells,
i.e. a few 10$^7$ years.
 This is unlikely since infall timescales are of order
a dynamical timescale which is typically a few 
10$^{8}$ yr.

While these arguments have been applied to NGC~3556 alone, the difficulty
with internal energy from starbursting applies to a number of galaxies, as
outlined in Sect.~1.  Infalling clouds are certainly plausible, but run into
difficulties for isolated galaxies such as NGC~3556 and NGC~3044.  
 Barring
new information coming to light on these galaxies, we therefore argue
that a new model should be considered.

\subsection{Outline of the Model}

As mentioned earlier, the duality of the expanding HI supershells in 
NGC~3556, together with their locations at the opposite edges
of the disk marked by steep density gradients, their Z-symmetric
deviations from spherical symmetry, as well as their anomalously 
large energy requirements (Sect.~2.3.1), together lead us to argue 
that the origin of these two supershells is linked to the jet phenomenon.
In Sect.~2 we have noted several manifestations of jet instabilities
giving rise to bubble-shaped features, or shells, in the form of optical 
nebulosities or nonthermal radio emission in normal spirals and radio galaxies. 
In the present context of HI supershells, we examine the possibility
of the supershell being inflated due to localised instabilities in 
the radio jets as they plough through the ISM of the disk of NGC~3556.
The present discussion should only be viewed as a feasibility check 
based on energy considerations, and not as a detailed quantitative model 
for the supershells. Even in the case of NGC~3556, the standard mechanisms 
for the supershells (Sect.~1) may well have contributed at some level.

According to our proposal, 
the two radio jets ejected close to the plane of the disk, during 
the active phase of the nucleus, undergo a rapid {\it flaring} as they encounter
the region of large scale density decline 
near the outskirts of the galactic disk. Consequently, 
both synchrotron plasma and the entrained thermal plasma deposited 
by the jets in these two regions get heated and the resulting high-pressure 
bubble of hot plasma undergoes a rapid expansion, sweeping the gas-rich 
ambient medium into the shape of the HI supershells. An idealization
to such a situation is the flaring of a jet crossing an ISM/IGM interface, 
which was first discussed in the context of radio galaxies analytically 
by Gopal-Krishna \& Wiita (1987), followed by two-dimensional (Norman et al.
 1988, Wiita et al. 1990, Zhang et al.
 1999) and three-dimensional (Loken et al. 1995, Hooda \& Wiita 
1996, 1998) numerical simulations. The simulations by Loken et al. showed 
that intermediate power radio jets associated with wide-angle-tail (WAT)
sources undergo a rapid flaring upon crossing the ISM/IGM density 
discontinuity where a moderately supersonic jet becomes subsonic.  
As shown by the numerical simulations, a density drop 
of a factor of just a few can cause such a flaring (e.g. Hooda \& Wiita 1996). 
Observational evidence for the jet flaring comes from Wide-Angle-Tail
(WAT) radio sources which are associated with the dominant members of
rich clusters of galaxies (e.g. O'Donoghue et al. 1993). Further
evidence is provided, e.g., by the recent radio/optical/X-ray study of 
the WAT radio galaxy 3C465 in the Abell cluster A2634 (Sakelliou \& 
Merrifield 1999). Particularly instructive for the present study is the 
case of the active disk galaxy IRAS 04210+0400; the two $\sim~10$ kpc 
long radio jets associated with this disk galaxy are seen to flare up 
near the opposite optical boundaries of the galaxy (Holloway et al 1996). 
%The `puffing' of the HI supershells at the edges of the disk in NGC~3556 
%may be associated with such a flaring of the putative radio jets/lobes 
%nearly $10^7~yr$ ago.
It may be noted that during their passage through the inner parts of
the galactic disk, the jets are likely to encounter large ISM
density fluctuations.  However, these are less prone to disrupt the jets
because of the higher jet thrust there, combined with the expected
small spatial scale of the density fluctuations compared with the
jets' cross-section.

Since the postulated radio jets in NGC~3556 are currently too weak for
detection, a point which we re-address below, we shall adopt here the 
jet parameters inferred for another massive disk galaxy, NGC~4258 
(Sect.~2.1), and apply them to known conditions in NGC~3556. 
The jet velocity, as measured from the emission line gas, is {\it at 
least} v$_\mathrm{j}$ = 2000 km s$^{-1}$ which, together with a density of 
$\rho_\mathrm{j}$ = 0.02 $m_\mathrm{p}$ g cm$^{-3}$, where 
$m_\mathrm{p}$ is the mass of the 
proton, correspond to a kinetic luminosity of L$_\mathrm{j}$ $\sim$ 10$^{42}$ 
erg s$^{-1}$ (Falcke \& Biermann 1999). Assuming that the ram pressure 
in the jet dominates over the internal thermal pressure, then across the 
shock front we require 
$\rho_\mathrm{j}\,{(v_\mathrm{j}-v_\mathrm{s})}^2\,=\,\rho_0\,v_\mathrm{s}^2$, where 
$\rho_0$ is the ambient gas density, and $v_\mathrm{s}$ is the shock velocity. 
Taking $\rho_0$ = 0.26 $m_\mathrm{p}$ g cm$^{-3}$ at the positions of the 
supershells (KI97), we obtain $v_\mathrm{s}$ = 435 km s$^{-1}$. This is slightly 
higher than the value of $\sim$ 300 km s$^{-1}$ estimated for NGC~4258. 
The shock velocity puts a minimum timescale on the duration of the jet 
in this model, since there must be sufficient time for the jet to
propagate out to the locations of the supershells at 12 kpc galactocentric 
radius.  Thus, the minimum duration of the jets is 2.7 $\times$ 10$^7$ yr
for the parameters in this illustration.
% which is close to the inferred kinematic ages of the two supershells (\S 999).

The mechanical power of the jet would then be $L_\mathrm{j}$ =
$1/2\,\rho_0 \pi\,R^2\,v_\mathrm{s}^3$ = 5 $\times$ 10$^{41}$ erg s$^{-1}$,
where the effective jet radius, R has a typical value of 1 kpc (e.g.
Cecil et al 1995). This is similar to the jet power estimated for the
well known spiral galaxy M~51, which is at the low end of the scale
of nuclear activity (cf. Ford et al 1985). 
In NGC~3556, the integrated mechanical luminosity of
the jet over its minimum lifetime is then
$\sim$ 4 $\times$ 10$^{56}$ erg.  
This lower limit
is adequate to account for the observed kinetic energy 
associated with the larger eastern supershell (Sect.~2.3.1).
%This is consistent with the typical radio luminosity of $\sim$ 10$^{40}$ 
%erg s$^{-1}$ for Seyfert galaxies, taking an efficiency of order of 
%a few percent. 
%Thus, over the minimum estimated duration of the jets in NGC~3556, the 
%integrated mechanical luminosity of the jet $\sim~4$ $\times$ 10$^{99}$ erg. 
%This lower limit is adequate to account for the observed kinetic energy 
 Here we have
assumed
%Thus, it would take only 2 $\times$ 10$^7$ yr to inject the 3 $\times$
%10$^{56}$ ergs needed to inflate the eastern supershell. This assumes that the
that the efficiency for converting the input energy into the kinetic 
energy of the shell is of order 1\% as usually taken for the multiple 
supernova model.  For a higher efficiency, the required jet power can
be lower.  
%Since the known kinematic age of the shells is 5 $\times$ 10$^7$ yr, 
%we only require an input of energy of this magnitude over only a 
%fraction of the lifetime of the shell. 
% Thus, the input energy from such a jet is indeed adequate to drive the 
%supershells.

%Assuming that these rather weak counterparts of the radio galaxy jets
%are supersonic, intermediate Mach number jets, the amount of ambient 
%entrained by such a jet (e.g. de Young 1986) will be $\dot M~\approx~
%\rho_0\,v_s\,\pi\,R^2$, where $R$ is the effective entrainment radius. 
%Taking $R$ = 0.9 kpc (Cecil et al. 1995) we find $\dot M$ = 7 M$_\odot$ 
%yr$^{-1}$ (cf. 3.6 M$_\odot$ yr$^{-1}$ for NGC~4258). Over the minimum 
%jet lifetime, the mass of the thermal gas entrained by each jet 
%would be 2 $\times$ 10$^8$ M$_\odot$. When deposited by the jet at
%the locations of their flaring, this hot gas can drive a jet-flare-driven
%superwind into the ambient neutral medium.
%[Note, however, that a reduction in $R$ to 100 pc (see Martin et al. 1989)
% would decrease this value by a factor of 80.]  

%The masses of the eastern and western supershells in NGC~3556 are 5.9
%$\times$ 10$^7$ M$_\odot$ and 2.3 $\times$ 10$^7$ M$_\odot$, respectively.
%This allows for the possibility that the supershells may have condensed
%out of the material which was originally swept by the jets.  Alternatively,
%the supershells may be inflated from pre-existing material around the 
%shock's location.

The expansion of a typical bubble in the ISM should then proceed similarly
to previously modelled scenarios, the difference being the source of
input energy.  Since the input energy  is much higher than conventional
supernovae, the bubbles are more likely to achieve blowout, providing a
natural conduit through which cosmic rays (including those supplied by the 
jets) can escape into the halo.  Thus, the presence of supershells is
expected to correlate with the existence of a nonthermal radio halo.

The locations of the anchor points of the two supershells in NGC~3556
suggest that the postulated blasting of each supershell would have 
occurred at a galactocentric distance which is just past the peak of 
the HI rotation curve (see Fig. 6 of KI97). The radially outward expansion 
of each supershell would then expose it to the velocity shear in the 
medium, exerting a side-ways wind pressure counter to the galactic rotation. 
Due to this, the half of the supershell towards the direction of the 
galactic rotation would be compressed and hence brightened, whilst the 
other half would be dragged out due to the velocity shear in the external 
medium and, consequently, dimmed. Such a deformation of the two supershells 
from spherical symmetry, in course of their expansion, would give rise to 
a Z-symmetry, which is consistent with the observations (Sect.~2.3.1; KI97).

The cooling time for the X-ray emitting sheaths around the jets of NGC~4258,
assuming unity filling factor,  is only 4 $\times$ 10$^6$ yr (Cecil et al.
1995) and therefore, once the jets have turned off, such a signature of
their existence would disappear rapidly.  The lifetime of the synchrotron
emitting particles in the jets is longer, typically from 10$^6$ to 10$^7$ yr.
For NGC~4258, it has been estimated to be between 1 to 5 $\times$ 10$^7$ yr
(Martin et al. 1989). However, shorter timescales are possible and will
depend on local magnetic field strength and spectral index. For instance,
the magnetic field in the north-east radio jet in NGC~1068, which extends
to $\sim$ 450 pc from the nucleus, is $\sim$ 5 $\times$ 10$^{-4}$ Gauss
and the spectral index is -1 (Wilson \& Ulvestad 1987).  This gives a
synchrotron lifetime of only 1.5 $\times$ 10$^5$ yr.  The particle lifetime
is expected to be shorter in shocks where the magnetic field is 
compressed.  Thus, we expect the radio jet in NGC~3556 to fade over a 
timescale less than the ages of the supershells.

Similar arguments apply to the radio core which is expected to decay
faster than the jets, once the nuclear activity has switched off, given 
the likelihood of a stronger magnetic field and flatter spectral index 
in the core.  Even if some radio emission from a core were to persist
after the activity
ceases, it may be below the rms noise level of the available maps.  
Chary \& Becklin (1997), for example, estimate the radio luminosity of 
the core of NGC~4258, which is known to have a supermassive central 
object of mass 3.6 $\times$ 10$^7$ M$_\odot$ and a VLBI jet, to be 
L$_\mathrm{rad}$ = 100 L$_\odot$.  The 2$\sigma$ noise level of the radio maps 
of Irwin et al. (2000) corresponds to a radio power of 6.8 $\times$ 
10$^{17}$ W Hz$^{-1}$. Integrating over 10$^{11}$ Hz, this yields a radio 
luminosity of 180 L$_\odot$. Thus, even if NGC~3556 harbours a radio core 
of the same luminosity as NGC~4258, it would not have been detected in the
observations mentioned above.

%The above example was meant to illustrate that the energy available in
%even low luminosity jets of the type known to exist in spiral galaxies,
%is adequate to form the supershells seen in NGC~3556.  It is possible,
%on a more detailed level, that other processes are important.

%For example, instabilites could inflate the bubbles instead of (or in
%addition to) the energy available from shocks.  Since the internal kinetic
%luminosity of the jet in NGC~4258 is L$_\mathrm{j}$ $\sim$ 10$^{42}$ erg s$^{-1}$
%(Falcke \& Biermann 1999) an instability of order one half of this value
%would be sufficient to drive the supershells at a level roughly equal to
%the shock scenario.

\section{Summary and Possible Implications}

In this study we have sketched a scenario whereby radio jets ejected during
an active nuclear phase in the life of a large spiral galaxy could inflate
out of the disk exceptionally large shells, i.e. supershells of neutral
hydrogen gas.  When applied to the Scd galaxy, NGC~3556, this model can 
account for each of the several features enumerated in Sect.~2.3.1.  
In the context of the HI supershells seen in this galaxy, this scheme
 appears to fare distinctly better than either of the two
standard models for supershells which
 invoke either a starburst induced superwind or an 
infall of external gas clouds on to the galaxy disk. 
  Thus, the jet-ISM interaction scenario for disk galaxies 
could effectively supplement the other two mechanisms already proposed 
for the supershell phenomenon, with a greater relevance whenever the 
shells are extraordinarily large as well as energetic and occur
at large galactocentric distances.

The energy requirement is no longer a major issue in the present model.
Even a modest energy input from a 10$^7$ M$_\odot$ compact object
at the galactic nucleus, such as that observed in NGC~4258, is more than
adequate to supply the input energy required for the supershells
of NGC~3556.  This model can therefore account for HI supershells
which are difficult to form via conventional processes involving
massive star formation, or those occurring in galaxies lacking potentially 
interacting companions.  Larger supershells (or parts thereof, given the 
probable development of instabilities) may also be predicted, since known 
central masses and/or accretion rates may well surpass those considered 
here.

Furthermore, the present model can account for the symmetric HI
features in galaxies.  For instance, it is more likely that bubbles will
be inflated by the propagating jets at galactocentric radii where the
ISM density has declined to sufficiently low values for blow-out. At the
same time, symmetry is not a generic outcome our model, since local 
shocks,
dictated by the interaction of the jet with local density perturbations,
can be important in determining where the bubbles will be inflated.
For example, known bubbles in the lobes of radio galaxies (Sect.~2.2) are not
always seen to be equidistant on either side of the nucleus.

In our sample illustration (Sect.~3.2), the jets must exist for long enough 
to reach large
galactocentric radii (3 $\times$ 10$^7$ yr) 
 but need to inject energy over
only a fraction of the lifetime of the supershells (2 $\times$ 10$^7$ yr,
or even less, if the efficiency exceeds 1\%).  The jets presumably switch
off thereafter, with the synchrotron emitting particles decaying in 
10$^6$ - 10$^7$ yr.  
It is difficult to predict whether jets and HI supershells should
co-exist in spirals.  Nuclear activity in galaxies is commonly believed to
be a transient phenomenon, lasting for $\sim~10^7 - 10^8$ years (e.g.
Eilek 1996, Scheuer 1995; Soltan 1982; Haehnelt \& Rees 1993; Ho et al
1997). If AGN activity
occurs on timescales longer than a few $\times$ 10$^7$ yr, this would suggest
that there might be disk galaxies in which both phenomena should be observed
at the same time.
However, such nuclear activity
timescales are likely to be more representative of radio galaxies;
AGN activity in spirals is probably shorter lived.  An important
next step is to consider what jet parameters are required to
reach large, or at least kpc-scale radii, in traversing the typical
ISM of a spiral galaxy.

If jets in spirals both recur and precess, it may be possible to inflate
bubbles over a variety of galactocentric radii and azimuthal angles.  However,
it is unlikely that the ``frothy" nature of the HI seen in galaxies such as
Holmberg II or the holes in M~31 could be produced by such jets directly
without seeing some lingering evidence for the jets as well.  
Also, the jet phenomenon cannot directly account for 
features which correlate with
the star forming disk, such as some known radio and H$\alpha$ halos. 
If there is a connection,
it is more likely to be indirect.  For example,
once the expansion of the postulated jet-blown supershells is halted due to
the gravitational pull of the host galaxy, their segments (i.e.  HI clumps)
would eventually shower back on to the galactic disk, giving rise to
secondary shells, bubbles and cavities in the HI component of the disk.
The impact could also trigger sporadic bursts of star formation,
especially in the outer disk, as seen, e.g., in gas-rich
low-surface-brightness galaxies (cf. O'Neil et al. 1997)
which are thought to
be the present-day remnants of quasars (e.g. Schombert 1998). Thus, 
 much after its cessation, the nuclear activity in gas-rich
galaxies may continue to influence the evolution of their disks.

%input{table}
%\section{The Model\label{sec:phases}}
%Between the release from a radio galaxy and the reappearance as a
%cluster radio relic the radio \c{plasmon} undergoes several different
%phases of expansion and contraction:
%\noindent
%{\bf Phase 0: Injection.} The radio galaxy is active and a large
%\noindent
%{\bf Phase 1: Expansion.} After the central engine of the radio galaxy

%\begin{figure}[t]
%\begin{center}
%\psfig{figure=syncA.ps,width=0.45 \textwidth,angle=0}
%\end{center}
%\caption[]{\label{fig:syncA} Radio spectrum of the radio cocoon in
%scenario A at the end of phases 0-4.}
%\end{figure}

%\begin{figure}[t]
%\begin{center}
%\psfig{figure=syncB.ps,width=0.45 \textwidth,angle=0}
%\end{center}
%\caption[]{\label{fig:syncB} Radio spectrum of the radio cocoon in
%scenario B at the end of phases 0-4. The luminosity at the end of
%phase 2 is too small in order to be displayed in this figure.}  
%\end{figure}

%\noindent

%\section{Discussion\label{sec:discussion}}

%Below, we summarize some merits and potential shortcoming of this model:\\

%\noindent
%{\bf Pros:}
%\begin{itemize}
%\item The observed connection of cluster radio relics to shock waves
%arises naturally in this model.
%\item The presence of the radio galaxy NGC 4789 and the morphological
%\end{itemize}

\acknowledgements 

It is a pleasure to thank Paul J. Wiita for valuable comments on the 
manuscript.  JI gratefully acknowledges a grant from the Natural Sciences 
and Engineering Research Council of Canada.

%\bibliography{tae}

\begin{thebibliography}{}

\bibitem {} Aoki K., Kosugi G., Wilson A. S., Yoshida M. 1999, ApJ 521, 565
\bibitem {} Axon D. J., Marconi A., Capetti A., et al.  1998, ApJ 496, 75
\bibitem {} Baum S. A., O'Dea C. P., de Koff S., et al. 1996, ApJ, 465, L5
\bibitem {} Blitz L., Spergel D. N., Teuben P. J., Hartmann D., Burton W. B.
1999, ApJ 514, 818
\bibitem {} Bloemen H., Duric N., Irwin J. A. 1993, in 23rd International Cosmic
Ray Conference, eds. D. A. Leahy, R. B. Hicks, D. Venkatesan (new Jersey, World Scientific),
p. 279
\bibitem {} Brinks E., Bajaja E. 1986, A\&A 169, 14 
\bibitem {} Cecil G., Wilson A. S., de Pree C. 1995, ApJ 440, 181
\bibitem {} Cecil G., de Pree C. G., Greenhill L. J., Moran J. M.,
Dopita M. A. 1998, AAS 193.0710 
\bibitem {} Chary R., Becklin E. E. 1997, ApJ 485, L75
\bibitem {} Chaves T., Irwin J. A., in Gas and Galaxy Evolution, ASP
Conf. Ser., ed. J. E. Hibbard, M. P. Rupen, J. H. van Gorkom, in press
\bibitem {} Chevalier, R. A. 1974, ApJ 188, 501
\bibitem {} Cox P., Downes D. 1996, ApJ 473, 219
\bibitem {} Dahlem M., Lisenfeld U., Golla G. 1995, ApJ 444, 119
\bibitem {} de Bruyn A. G. 1977, A\&A, 58 221
\bibitem {} de Bruyn A. G.,  Hummel E. 1979, A\&A 73, 196
\bibitem {} Dreher J. W.,  Feigelson E. D. 1984, Nature 308, 43
\bibitem {} Efremov Y. N., Elmegreen B. G.,  Hodge P. W. 1998, ApJ 501, L163
\bibitem {} Eilek J. A. 1996, in Energy Transport in Radio Galaxies
and Quasars, ASP Conf. Ser. 100, ed. P.E. Hardee, A.H. Bridel, J.A. Zensus, 
(San Francisco: Astr. Soc. Pacific), 281
\bibitem {} Elmouttie M., Haynes R. F., Jones K. L., Sadler E. M., Ehle M.
1998, MNRAS 297, 1202
\bibitem {} Falcke H., Biermann P. L. 1999, A\&A 342, 49
\bibitem {} Ford H. C., Crane P. C., Jacoby G. H., Lawrie D. G., van der
Hulst J M. 1985, ApJ 293, 132
\bibitem {} Genzel R., Eckart A., Ott T., Eisenhauer F. 1997,
MNRAS 291, 219
\bibitem {} Gopal-Krishna, Wiita P. J. 1987, MNRAS 226, 531
\bibitem {} Haehnelt M., Rees M. 1993, MNRAS 263, 168
\bibitem {} Heiles C. 1979, ApJ 229, 533
\bibitem {} Heiles C. 1984, ApJS 55, 585
\bibitem {} Herrnstein J. R., Moran J. M., Greenhill L. J., et al.
 1997, ApJ 475, L17, Erratum, ApJ, 482, L113
\bibitem {} Ho L. C., Filippenko A. V., Sargent W. L. W., 1997,
ApJ 487, 568
\bibitem {} Holloway A. J., Steffen W., Pedlar A., et al. 1996, MNRAS 279, 171
\bibitem {} Hooda J. S., Wiita P. J. 1996, ApJ 470, 211
\bibitem {} Hooda J. S., Wiita P. J. 1998, ApJ 493, 81
\bibitem {} Hummel E., van Gorkom J. H., Kotanyi C. G. 1983, ApJ 267, L5
\bibitem {} Irwin J. A. 1994, ApJ 429, 618
\bibitem {} Irwin J. A., Seaquist E. R. 1990, ApJ 353, 469
\bibitem {} Irwin J. A., English J., Sorathia B. 1999, AJ 117, 2102
\bibitem {} Irwin J. A., Saikia D. J., English J. 2000, AJ 119, 1592
\bibitem {} King D., Irwin J. A. 1997, New Astr. 2, 251 (KI97)
\bibitem {} Kinney A. L.,  2000, ApJ (July 2000 issue)
\bibitem {} Kormendy J. Richstone D. 1995, ARAA 33, 581
\bibitem {} Kukula M. J., Pedlar A., Baum S. A., O'Dea, C. P. 1995, MNRAS 276, 1262
\bibitem {} Lee S.-W., Irwin J. A. 1997, ApJ 490, 247 (LI97)
\bibitem {} Lee S.-W., Irwin J. A., Dettmar R.-J., et al. 2000,
in prep.
\bibitem {} Lehnert M. D., Heckman T. M. 1996, ApJ 462, 651
\bibitem {} Loeb A., Perna R., 1998, ApJ 503, L35
\bibitem {} Loken C., Roettiger K., Burns J. O., Norman M. 1995,
ApJ 445, 80
\bibitem {} Martin P., Roy J.-R., Noreau L., Lo K. Y. 1989, ApJ 345, 707
\bibitem {} Meurer G. R., Heckman T. M., Leitherer C., et al. 1995, AJ 110, 2665
\bibitem {} Moody J. W., Roming P. W. A., Joner M. D., et al. 1995, AJ 110, 2088
\bibitem {} Morganti R., Killeen N. E. B., Ekers R. D., Oosterloo T. A. 1999,
MNRAS 307, 750
\bibitem {} Morrison P., Sadun A. 1996, MNRAS 278, 265
\bibitem {} Nagar N. M., Wilson A. S. 1999, ApJ 516, 97
\bibitem {} Norman M. L., Burns J. O., Sulkanen M. 1988, Nature 335, 146
\bibitem {} Norman C. A., Ikeuchi S. 1989, ApJ 345, 372
\bibitem {} O'Donoghue A. A., Eilek J. A., Jones J. M. 1993, ApJ 408, 428
\bibitem {} O'Neil K., Bothun G. D., Schombert J., Cornell M. E., Impey C. D. 1997,
AJ 114, 2448
\bibitem {} Puche D., Westpfahl D., Brinks E.,  Roy J.-R. 1992,
AJ 103, 1841
\bibitem {} Rand R. J. 1997, in The Interstellar Medium in Galaxies, ed. J. M.
van der Hulst (Dordrecht: Kluwer), 105
\bibitem {} Rand R. J., Stone  J. M. 1996, AJ 111, 190
\bibitem {} Rand R. J., van der Hulst J. M. 1993, AJ 105, 2098, Erratum,
AJ, 107, 392
\bibitem {} Rhode K. L., Salzer J. J., Westpfahl D. J., Radice L. A. 1999,
AJ 118, 323
\bibitem {} Sakelliou I., Merrifield M. R. 1999, MNRAS 305, 417
\bibitem {} Scheuer P. A. G. 1995, MNRAS 277, 331
\bibitem {} Schombert J. 1998, AJ 116, 1650
\bibitem {} Soltan A. 1982, MNRAS 200 115
\bibitem {} Tenorio-Tagle G., Bodenheimer P., 1988, ARA\&A 26, 145
\bibitem {} Ulvestad J. S., Wilson A. S. 1984a, ApJ 278, 544
\bibitem {} Ulvestad J. S., Wilson A. S. 1984b, ApJ 285, 439
\bibitem {} Ulvestad J. S., Wrobel J. M., Roy A. L., et al. 1999, ApJ 517, L81
\bibitem {} van Breugel W., Fomalont E. B. 1984, ApJ 282, L55
\bibitem {} van der Marel R. P., 1999, AJ 117, 744
\bibitem {} van der Hulst T., Sancisi R. 1988, AJ 95, 1354
\bibitem {} Wiita P. J., Rosen A., Norman M. L. 1990, ApJ 350, 545
\bibitem {} Wilson A. S., Ulvestad J. S. 1987, ApJ 319, 105
\bibitem {} Zhang H.-M., Koide S., Sakai J.-I. 1999, PASJ 51, 449
\end{thebibliography}
%\bibliographystyle{aabib99}
\end{document}